\documentclass[nofootinbib, superscriptaddress, twocolumn,secnumarabic,amssymb, nobibnotes, aps, prl, longbibliography]{revtex4-1}

\newcommand{\epi}{\pi}
\newcommand{\ket}[1]{|#1\rangle}
\newcommand{\bra}[1]{\langle#1|}
\usepackage{siunitx}
\usepackage{graphicx}
\usepackage{amsmath}
\usepackage{bm}
\usepackage{color}
\usepackage{isomath}
\usepackage{epsfig}
\usepackage[normalem]{ulem}

\begin{document}

\title{High-fidelity control and entanglement of Rydberg atom qubits}

\newcommand{\Caltech}{Division of Physics, Mathematics and Astronomy, California Institute of Technology, Pasadena, CA 91125, USA}
\newcommand{\Harvard}{Department of Physics, Harvard University, Cambridge, MA 02138,
USA}
\newcommand{\mitaddress}{Department of Physics and Research Laboratory of Electronics, Massachusetts Institute of Technology,
Cambridge, MA 02139, USA}

\newcommand{\LKB}{Laboratoire Kastler Brossel, ENS-PSL Research University, CNRS, Sorbonne Universit\'e, Coll\`ege de France, 24 rue Lhomond, 75005 Paris, France }

\author{Harry Levine}
\address{\Harvard}

\author{Alexander Keesling}
\address{\Harvard}

\author{Ahmed Omran}
\address{\Harvard}

\author{Hannes Bernien}
\address{\Harvard}

\author{Sylvain Schwartz}
\address{\LKB}

\author{Alexander S. Zibrov}
\address{\Harvard}

\author{Manuel Endres}
\address{\Caltech}

\author{Markus Greiner}
\address{\Harvard}

\author{Vladan Vuleti\'c}
\address{\mitaddress}

\author{Mikhail D. Lukin}
\address{\Harvard}

\begin{abstract}
  Individual neutral atoms excited to Rydberg states are a promising platform for quantum simulation and quantum information processing.  However, experimental progress to date has been limited by short coherence times and relatively low gate fidelities associated with such Rydberg excitations. We report progress towards high-fidelity quantum control of Rydberg atom qubits. Enabled by a reduction in laser phase noise, our approach yields a significant improvement in coherence properties of individual qubits.  We further show that this high-fidelity control extends to the multi-particle case by preparing a two-atom entangled state with a fidelity exceeding  0.97(3), and extending its lifetime with a two-atom dynamical decoupling protocol. These advances open up new prospects for scalable quantum simulation and quantum computation with neutral atoms.
\end{abstract}

\maketitle

Neutral atoms are attractive building blocks for large-scale quantum systems. They can be well isolated from the environment, enabling long-lived quantum memories. Initialization, control, and read-out of their internal and motional states is accomplished by resonance methods developed over the past four decades \cite{WeissSaffman2017}. Recent experiments demonstrated that arrays with a large number of identical atoms can be rapidly assembled while maintaining single-atom optical control \cite{AntoineAssembly2016, AtomArrayScience2016, AhnAssembly2016}. These bottom-up approaches are  complementary to the methods involving optical lattices loaded with ultracold atoms prepared via evaporative cooling \cite{OpticalLattices2017}, and generally result in atom separations of several micrometers.  In order to utilize these arrays for quantum simulation and quantum information processing, it is necessary to introduce controllable interactions between the atoms.  This can be achieved  by coherent coupling to highly excited Rydberg states, which exhibit strong, long-range interactions \cite{AntoineReview2016}. Over the past decade, this approach has emerged  as a powerful platform for many applications, including fast multi-qubit quantum gates \cite{Jaksch2000, SaffmanReview2016, AntoineEntanglementUsingBlockade2010, SaffmanCNOT2010, Saffman2DCNOT2015, Biedermann2016, ZhanDifferentIsotopes2017}, quantum simulations of Ising-type spin models with up to 250 spins \cite{MunichCrystals2015, AntoineTunableNature2016, MunichDressing2016, AtomArrayNature2017, AntoineAdiabatic2017, WaseemAdiabatic2017, AhnThermalization2018}, and the study of collective behavior in mesoscopic ensembles \cite{Weidemuller2004, Pfau2007, Kuzmich2012, Ott2015, NonlinearQuantumOpticsReview2016}. Despite these impressive demonstrations,  experimental progress to date has been limited by short coherence times and relatively low gate fidelities associated with such Rydberg excitations \cite{SaffmanReview2016}. This imperfect coherence limits the quality of quantum simulations, and especially dims the prospects for  neutral atom quantum information processing. The limited coherence becomes apparent even at the level of single isolated atomic qubits \cite{AntoineCoherence2018}.

This Letter reports the experimental realization of high-fidelity quantum control of Rydberg atom qubits. We show that by reducing laser phase noise, a significant improvement in the coherence properties of individual qubits can be achieved, consistent with recent theoretical analysis \cite{AntoineCoherence2018}.
We further demonstrate that this high-fidelity control extends to the multi-particle case by preparing a two-atom entangled state with a fidelity exceeding  0.97(3).
Finally, we extend the lifetime of the prepared Bell state  with a novel two-atom dynamical decoupling protocol.

Our experimental setup has been described in detail previously \cite{AtomArrayScience2016, AtomArrayNature2017}. We deterministically prepare individual cold Rubidium-87 atoms in optical tweezers at programmable positions in one dimension. The atoms are initialized in a Zeeman sublevel $\ket{g} = \ket{5S_{1/2}, F=2, m_F=-2}$ of the ground state via optical pumping in a 1.5 G magnetic field. We then rapidly switch off the tweezer potentials, and apply a laser field to couple the atoms to the  Rydberg state $\ket{r} = \ket{70S, J=1/2, m_J = -1/2}$. After the laser pulse of typical duration 3-8~$\mu$s, we restore the tweezer potentials. Atoms that are in the ground state are recaptured by the tweezers, whereas those left in the Rydberg state are pushed away by the tweezer beams \cite{AntoineCoherence2018}. This simple detection method has Rydberg state detection fidelity $f_r = 0.96(1)$ and ground state detection fidelity $f_g$ ranging from $0.955(5)$ to $0.990(2)$, depending on the trap-off time \cite{Supplement}. 

\begin{figure}
  \includegraphics{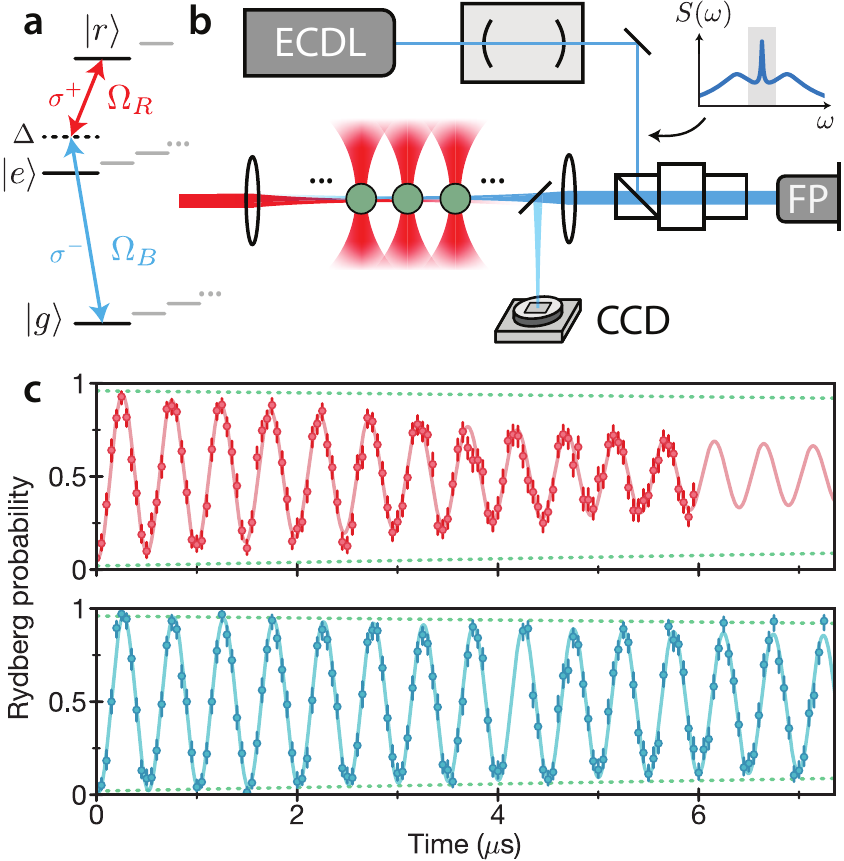}
  \caption{Experimental setup and single-atom Rabi oscillations. (a) The ground state $\ket{g} = \ket{5S_{1/2}, F=2, m_F=-2}$ is coupled to $\ket{r} = \ket{70S, J=1/2, m_J=-1/2}$  via the intermediate state $\ket{e} = \ket{6P_{3/2}, F=3, m_F=-3}$. (b) The lasers are locked to a reference cavity whose  narrow transmission window (shaded region in inset) suppresses high-frequency phase noise. This transmitted light is used to injection lock a Fabry-Perot (FP) laser diode. The  laser diode output is focused onto the array of atoms trapped in optical tweezers, with a small pickoff onto a reference CCD camera used for alignment. (c) Resonant two-photon coupling induces Rabi oscillations between $\ket{g}$ and $\ket{r}$. The upper plot is a typical measurement from the previous setup used in \cite{AtomArrayNature2017}. The lower plot shows typical results with the new setup, with a fitted coherence time of $27(4)\mu$s. Each data point is calculated from 50-100 repeated measurements, averaged over two identically-coupled atoms separated by $23~\mu$m such that they are non-interacting. Error bars are 68\% confidence intervals. Solid lines are fits to experimental data, and the dotted lines indicate the expected contrast from the numerical model.}
\end{figure}

In our experiments, the Rydberg states are excited via  a two-photon transition. A 420 nm laser is blue detuned by $\Delta$ from the transition from $\ket{g}$ to $\ket{e} = \ket{6P_{3/2}, F=3, m_F=-3}$. A second laser field at $1013~$nm couples $\ket{e}$ to $\ket{r}$. The two lasers are polarized to drive $\sigma^-$ and $\sigma^+$ transitions, respectively, such that only a single intermediate sublevel and Rydberg state can be coupled, avoiding the population of additional levels and associated dephasing (see Fig. 1(a)). 

The two lasers (external-cavity diode lasers from MogLabs) are frequency stabilized by a Pound-Drever-Hall (PDH) lock to an ultra-low expansion reference cavity (StableLasers). The PDH lock strongly suppresses laser noise at frequencies below the effective bandwidth of the lock, resulting in narrow linewidths of $< 1~$kHz, as estimated from in-loop noise. However, noise above the lock bandwidth cannot be suppressed, and can be amplified at high locking gain. This results in broad peaks in phase noise around $\sim 2\pi \times 1$ MHz (see inset of Fig. 1(b)). This high-frequency phase noise has been reported as a known coherence limitation in Rydberg experiments \cite{AntoineCoherence2018} and experiments with trapped ions \cite{SpectralFilteringIsrael2015, SpectralFilteringETH2015}, and has also been studied in the context of atomic clocks \cite{SpectralFilteringJILAClocks2013}. To suppress this phase noise, we follow the approach of \cite{SpectralFilteringHald2005, SpectralFilteringSterr2008, SpectralFilteringIsrael2015, SpectralFilteringETH2015} in which the reference cavity is used as a spectral filter. In particular, the transmission function of the cavity is a Lorentzian with a full-width at half maximum of $\Gamma \sim 2\pi\times 500~$kHz (corresponding to a finesse of $F \sim 3000$). When the laser is locked, its narrow linewidth carrier component is transmitted through the cavity, whereas the high-frequency noise at $2\pi \times 1$~MHz is suppressed by a factor of $\gtrsim 4$. To amplify this light at both 420 and 1013 nm, we split the two colors and use each beam to injection lock a separate laser diode (1013 nm from Toptica, 420 nm from TopGaN), which inherits the same spectral properties. This  amplifies the spectrally pure transmitted light to 5 mW of 420 nm and 50 mW of 1013 nm light. While the 420 nm power is sufficient to drive the blue transition directly, the 1013 nm is further amplified by a tapered amplifier (MogLabs). 

We focus both lasers onto the atom array in a counter-propagating configuration to minimize Doppler shifts due to finite atomic temperature. The 420 (1013) nm laser is focused to a waist of 20 (30) $\mu$m. We achieve single-photon Rabi frequencies of $\Omega_B \simeq 2\pi \times 60~\text{MHz} \ (\Omega_R \simeq 2\pi \times 40~\text{MHz})$. At our intermediate detuning of $\Delta \simeq 2\pi \times 600~\text{MHz}$, this leads to a two-photon Rabi frequency of $\Omega = \Omega_B \Omega_R / (2\Delta) \simeq 2\pi \times 2~\text{MHz}$. Each beam is power-stabilized to $< 1\%$ by an acousto-optic modulator that is also used for fast ($\sim 20$ ns) switching. To minimize sensitivity to pointing fluctuations, we ensure well-centered alignment onto the atoms using a reference camera (depicted in Fig. 1(b)) and an automatic beam alignment procedure \cite{Supplement}.

With these technical improvements in place, we measure long-lived Rabi oscillations with a $1/e$ lifetime of $\tau = 27(4)~\mu$s, to be compared with a typical $\lesssim7~\mu$s lifetime in previous experiments \cite{AtomArrayNature2017} (see Fig. 1(c)). Importantly, we observe excellent agreement between these new measurements and a simple numerical model for our single-atom system, indicated by dotted lines in Fig. 1(c). The numerical model has no free parameters and accounts only for the effects of random Doppler shifts at finite atomic temperature, off-resonant scattering from the intermediate state $\ket{e}$, and the finite lifetime of the Rydberg state $\ket{r}$ (see \cite{Supplement} for details). The results from the numerical model are additionally scaled to account for detection fidelity.

\begin{figure}
  \includegraphics{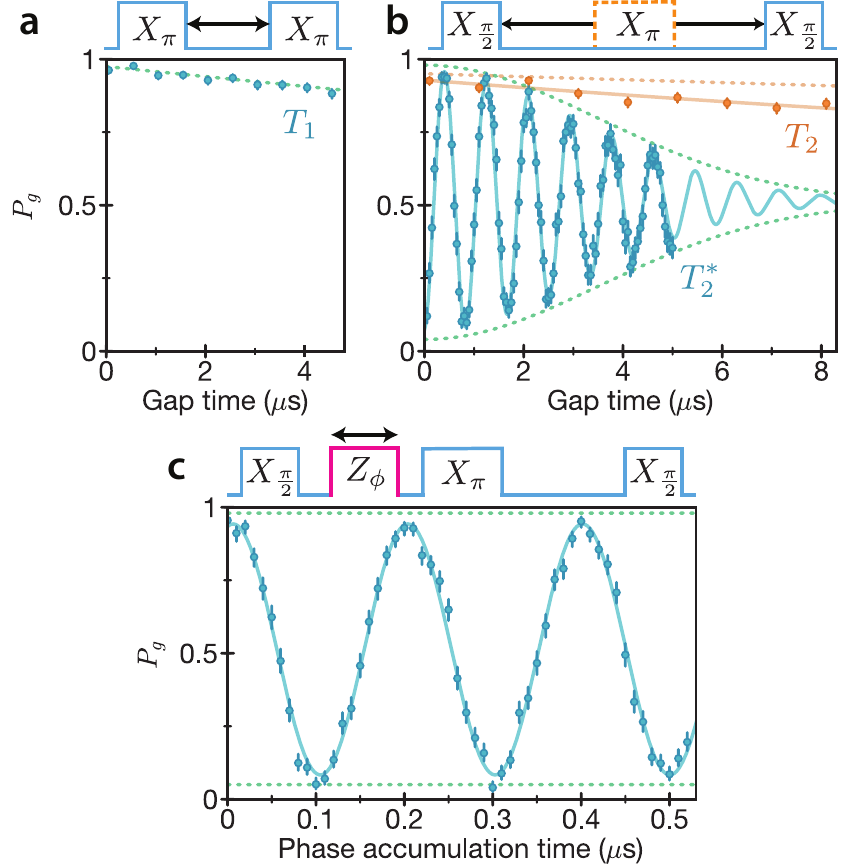}
  \caption{Characterization of single-atom coherence and phase control. (a) The lifetime of $\ket{r}$ is measured by exciting from $\ket{g}$ to $\ket{r}$ with a $\pi$-pulse, and then  de-exciting after a variable delay.  The probability to end in $\ket{g}$ (denoted $P_g$) decays with an extracted lifetime of $T_1 = 51(6)~\mu$s. (b) A Ramsey experiment (blue) shows Gaussian decay with a $1/e$  lifetime of $T_2^* = 4.5(1)~\mu$s, limited by thermal Doppler shifts. Inserting an additional $\pi$-pulse (orange) between the $\pi/2$-pulses cancels the effect of the Doppler shifts and results in a substantially longer coherence lifetime of $T_2=32(6)~\mu$s (fitted to an exponential decay down to 0.5). (c) A single-atom phase gate  is implemented by applying an independent 809 nm laser which induces a light shift $\delta = 2\pi \times 5~$MHz on the ground state for time $t$, resulting in an accumulated dynamical phase $\phi = \delta t$. The gate is embedded in a spin-echo sequence to cancel Doppler shifts. In each measurement shown here, the 1013 nm laser remains on for the entire pulse sequence, while the 420 nm laser is  pulsed according to the sequence shown above each plot. Each data point is calculated from $200-500$ repeated measurements on single atoms, with error bars denoting $68\%$ confidence intervals. In (b) and (c), the solid lines are fits to experimental data. Dotted lines show the contrast expected from the numerical model, including finite detection fidelity.}
\end{figure}

Next, we characterize the coherence of single atoms and demonstrate single-qubit control. To begin, we experimentally measure the lifetime of the Rydberg state in Fig. 2(a). The measured $T_1 = T_{r \to g} = 51(6)~\mu$s is consistent with the $146~\mu$s Rydberg state lifetime \cite{RydbergProperties2009} when combined with the $\sim 80~\mu$s timescale for off-resonant scattering of the 1013 nm laser from $\ket{e}$. A  Ramsey experiment shows Gaussian decay that is well-explained by thermal Doppler shifts (see Fig. 2(b)). At $10~\mu$K, the random atomic velocity in each shot of the experiment appears as a random detuning $\delta^D$ from a Gaussian distribution of  width $2\pi \times 43.5$~kHz, resulting in dephasing as $\ket{\psi} \to \frac{1}{\sqrt{2}}(\ket{g} + e^{i\delta^D t} \ket{r})$. However, since the random Doppler shift is constant over the duration of each pulse sequence,  its effect can be eliminated via a spin-echo sequence (orange in Fig. 2(b)). Note that the spin-echo measurements display some small deviations from the  numerical simulations, indicating the presence of an additional dephasing channel.  Assuming an exponential decay, we measure a fitted $T_2 = 32(6)~\mu$s and extract a pure dephasing time $T_\phi = (1/T_2 - 1/(2T_{r \to g}))^{-1} = 47(13)~\mu$s. We hypothesize that this dephasing may result from  residual laser phase noise. 

Finally, we demonstrate a single-atom phase gate by applying an independent focused laser that shifts the energy of the ground state $\ket{g}$ by 5~MHz. By controlling the duration of the applied laser pulse, we impart a controlled dynamical phase on $\ket{g}$ relative to $\ket{r}$. The  contrast of the resulting phase gate (embedded in a spin-echo sequence) is close to the limit imposed by detection and spin-echo fidelity.

We next turn to two-atom control. To this end, we position two atoms at a separation of $5.7~\mu$m, at which the Rydberg-Rydberg interaction is $U/\hbar=2\pi \times 30~\text{MHz} \gg \Omega=2\pi \times 2~$MHz. In this so-called Rydberg blockade regime, the laser field globally couples both atoms from $\ket{gg}$ to the symmetric\footnote{Here the excited states $|r\rangle$ are defined in the rotating frame to incorporate the spatial phase factors $e^{ikx}$, as discussed in \cite{Supplement}.} state $\ket{W} = \frac{1}{\sqrt{2}}(\ket{gr} + \ket{rg})$ at an enhanced Rabi frequency of $\sqrt{2}\Omega$ (see Fig. 3(a)). The measured probabilities for the states $\ket{gg}, \ket{gr},  \ket{rg}$, and $\ket{rr}$ (denoted by $P_{gg}, P_{gr}, P_{rg}$, and $P_{rr}$, respectively) show that indeed no population enters the doubly-excited state $(P_{rr} < 0.02$, consistent with only detection error). Instead, there are oscillations between the manifold of zero excitations and the manifold of one excitation with a fitted frequency of $2\pi \times 2.83~\text{MHz} \approx \sqrt{2} \Omega$ (see Fig. 3(b)).

\begin{figure}
  \includegraphics{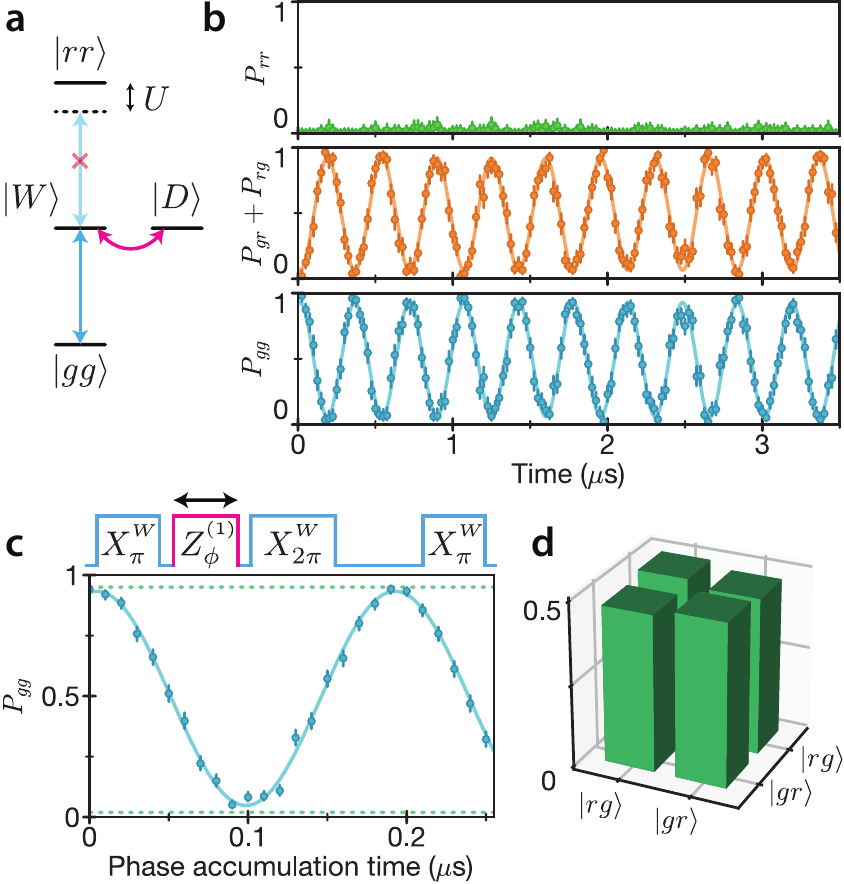}
  \caption{Coherent control and entanglement generation with two atoms. (a) The level structure for two nearby atoms features a doubly excited state $\ket{rr}$ which is shifted by the  interaction energy $U \gg \hbar \Omega$. In this Rydberg blockade regime, the laser field only  couples $\ket{gg}$ to $\ket{W}$. The symmetric and antisymmetric states $\ket{W}, \ket{D} = \frac{1}{\sqrt{2}}(\ket{gr} \pm \ket{rg})$ can be coupled by a local phase gate on one atom (pink arrow). (b) After driving both atoms on resonance for variable time, we measure the probability of the resulting two-atom states. Population  oscillates from $\ket{gg}$ to $\ket{W}$ at the enhanced Rabi frequency $\sqrt{2} \Omega$. (c) We measure the entanglement fidelity of the two atoms after a resonant $\pi$-pulse in the blockade regime. A local phase gate $Z_{\phi}^{(1)}$ rotates $\ket{W}$ into $\ket{D}$, which is detected by a subsequent $\pi$-pulse. The fitted contrast $0.88(2)$ measures the off-diagonal density matrix elements. The phase gate is implemented by an off-resonant laser focused onto one atom, with a crosstalk of $< 2\%$. The measurement is embedded in a spin-echo sequence to cancel dephasing from thermal Doppler shifts. (d) The four components of the density matrix correspond to an entangled state with fidelity $\mathcal{F} = 0.97(3)$ (corrected for detection error). Each data point in (b) and (c) is calculated from $\sim 50$ and $\sim 250$ repeated measurements, respectively, with error bars marking 68\% confidence intervals. The solid lines shown in (b) and (c) are fits to experimental data. Dotted lines in (c) mark the limits of detection fidelity.}
\end{figure}

These collective Rabi oscillations can be used to directly prepare the maximally entangled Bell state $\ket{W}$ by applying a $\pi$-pulse at the enhanced Rabi frequency (denoted by $X_\pi^W$). To determine 
 the fidelity of this experimentally prepared entangled state, given by $\mathcal{F} = \langle W | \rho | W \rangle$, we 
 express it in terms of diagonal and off-diagonal matrix elements of the density operator $\rho$:
 \begin{equation}
   \label{fidelity_definition}
\mathcal{F} = \frac{1}{2}(\rho_{gr,gr} + \rho_{rg, rg}) + \frac{1}{2}(\rho_{gr,rg} + \rho_{rg,gr})
\end{equation}
 where $\rho_{\alpha \beta, \gamma \delta} = \langle \alpha \beta | \rho | \gamma \delta \rangle$ for $\alpha, \beta, \gamma, \delta \in \{ g, r \}$.
 The diagonal elements can be directly measured by applying a $\epi$-pulse and then measuring the populations. The results closely match those of a perfect $\ket{W}$ state after accounting for state detection errors, with $\rho_{gr,gr} + \rho_{rg,rg} = 0.94(1)$, relative to a maximum possible value of $0.95(1)$.

 To measure the off-diagonal elements of the density matrix, we make use of the single-atom phase gate $Z_\phi^{(1)}$ demonstrated in  Fig. 2(c), which introduces a variable phase on one atom (as demonstrated in \cite{AntoineParity2014}). Specifically, a local beam adds a light shift $\delta$ to $\ket{gr}$ but not to $\ket{rg}$, such that $\ket{W} \to \frac{1}{\sqrt{2}}(e^{i\delta t}\ket{gr} + \ket{rg})$.
This phase accumulation rotates $\ket{W}$ into the orthogonal dark state $\ket{D} = \frac{1}{\sqrt{2}}(\ket{gr} - \ket{rg})$ according to:
\begin{equation}
  \label{w_state_to_d_state}
  \ket{W} \to \cos(\delta t/2)\ket{W} + i\sin(\delta t/2)\ket{D}
\end{equation}
Since $\ket{D}$ is uncoupled by the laser field, a subsequent $\epi$-pulse maps only the population of $\ket{W}$ back to $\ket{gg}$. The probability of the system to end in $\ket{gg}$ therefore depends on the phase accumulation time as $P_{gg}(t) = A\cos^2(\delta t/2)$. Here, the amplitude of the oscillation $A$ precisely measures the off-diagonal matrix elements $\rho_{gr,rg} = \rho_{rg,gr}$ (see \cite{Supplement} for derivation). Note that in order to mitigate sensitivity to random Doppler shifts, we embed this entire sequence in a spin-echo protocol (see Fig. 3(c)). The resulting contrast is $A = 0.88(2) = 2\rho_{gr,rg} = 2\rho_{rg,gr}$. Combining these values with the diagonal matrix elements, we have directly measured entanglement fidelity of $\mathcal{F} = 0.91(2)$. The maximum measurable fidelity given our state detection error rates would be $0.94(2)$, so after correcting for imperfect detection, we find that the entangled Bell state was created with fidelity of $\mathcal{F} = 0.97(3)$. We note that this fidelity includes errors introduced during the pulses that follow the initial $\epi$-pulse, and therefore constitutes a lower bound on the true fidelity.

\begin{figure}
  \includegraphics{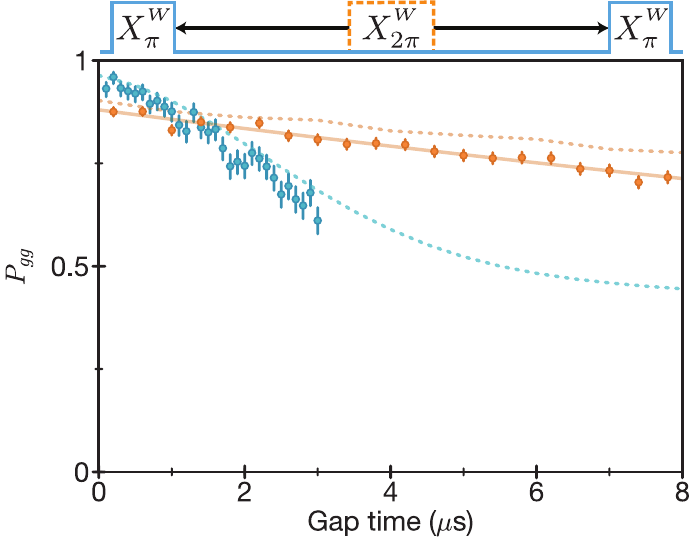}
  \caption{Extension of entangled-state lifetime via dynamical decoupling. We measure the lifetime of $\ket{W}$ by exciting $\ket{gg}$ to $\ket{W}$ and then de-exciting after a variable time (blue). The lifetime is limited by dephasing from random Doppler shifts. Inserting an additional $2\pi$-pulse (orange) in the blockade regime swaps the populations of $\ket{gr}$ and $\ket{rg}$ to refocus the random phase accumulation, extending the lifetime to $\sim 36~\mu$s (fitted to an exponential decay, shown as the solid orange line). The initial offset in each curve is set by the ground state detection fidelity associated with the given trap-off time. All data points are calculated from 30-100 repeated measurements, averaging over nine independent identically-coupled atom pairs, with error bars indicating 68\% confidence intervals. Dotted lines underneath data show predictions from the numerical model, including detection error. }
\end{figure}

Finally, we explore the  lifetime of the entangled state by exciting $\ket{W}$ with a $\epi$-pulse and then de-exciting after a variable delay (see Fig. 4). The decay in contrast is in good agreement with numerical predictions associated with random Doppler shifts. 
In particular, the two components $\ket{gr}$ and $\ket{rg}$ of the $\ket{W}$ state dephase as 
$
 \ket{W} \to \frac{1}{\sqrt{2}}(e^{i\delta_2^D t}\ket{gr} + e^{i\delta_1^D t} \ket{rg}),
$
where $\delta_i^D$ is the two-photon Doppler shift on atom $i$.

We extend the lifetime of the two-atom entangled state with a  many-body echo sequence. After the $\ket{W}$ state has evolved for time $T$, we apply a $2\epi$-pulse to the two-atom system. In the Rydberg blockade regime, such a pulse swaps the populations of $\ket{gr}$ and $\ket{rg}$. After again  evolving for time $T$, the total accumulated Doppler shifts are the same for each part of the two-atom wavefunction, and therefore do not affect the final $\ket{W}$ state fidelity. Indeed, Figure 4 shows that its lifetime is extended far beyond the Doppler-limited decay to $T_2^W = 36(2)~\mu$s. As in the single atom case, we extract a pure dephasing timescale  $T_{\phi}^W = (1/T_2^W - 1/T_{r \to g})^{-1}  > 100~\mu$s.

Remarkably,  the Bell state dephasing time $T_{\phi}^W > 100~\mu$s is significantly longer than the single atom dephasing time $T_\phi = 47(13)~\mu$s. This can be understood by noting that the states $\ket{gr}$ and $\ket{rg}$ form a decoherence-free subspace that is insensitive to global perturbations  such as laser phase and intensity fluctuations that couple identically to both atoms \cite{DFSWineland2001, DFSRoos2004}. In contrast, a single atom in a superposition $\ket{\psi} = \frac{1}{\sqrt{2}} (\ket{g} + \ket{r})$ is sensitive to both the laser phase and the  laser intensity. These measurements provide further indications that  laser noise is not completely eliminated in our experiment. 

Our measurements establish Rydberg atom qubits as a competitive platform for high-fidelity quantum simulation and computation. The techniques demonstrated in this Letter are of immediate importance to ongoing experiments using neutral atom arrays. Furthermore, the demonstrated fidelities can be further improved by increasing laser intensities and operating at larger detunings from the intermediate state, thereby reducing the deleterious effect of off-resonant scattering, or by using a direct single-photon transition. In addition, sideband cooling of atoms in tweezers \cite{AdamSidebandCooling2012, JeffSidebandCooling2013} can dramatically decrease the magnitude of  Doppler shifts, while  low-noise laser sources such as Titanium-Sapphire lasers or diode lasers filtered by higher-finesse cavities will further eliminate errors caused by phase noise. Advanced control techniques, such as laser pulse shaping, can also be utilized to reach higher fidelities \cite{SaffmanDRAG2016}. Finally, state detection fidelities, the major source of imperfections in the present work, can be improved by field ionization of Rydberg atoms or by mapping Rydberg states to separate ground state levels.

We acknowledge  A. Browaeys, M. Saffman, G. Biedermann and their groups for many fruitful discussions during the ITAMP workshop, which stimulated this study. We also thank J. Ye and T. Lahaye for many useful discussions and suggestions. This work was supported by NSF, CUA, ARO, MURI, AFOSR, and the Vannevar Bush Faculty Fellowship. H.L. acknowledges support from the National Defense Science and Engineering Graduate (NDSEG) fellowship. A.O. acknowledges support by a research fellowship from the German Research Foundation (DFG). S.S. acknowledges funding from the European Union under the Marie Sk{\l}odowska Curie Individual Fellowship Programme H2020-MSCA-IF-2014 (project number 658253).

%

\onecolumngrid

\section{Supplementary Information}

\subsection{Characterizing state detection fidelity}
The Rydberg pulse that couples the ground state $\ket{g}$ to the Rydberg state $\ket{r}$ is applied within a time window $\tau$ during which the optical tweezers are briefly turned off. Atoms that are left in $\ket{g}$ at the end of the pulse are recaptured by the optical tweezers, whereas those left in $\ket{r}$ are repelled by the tweezers and lost. We characterize this detection scheme by the ground state detection fidelity $f_g$ (the probability for an atom left in $\ket{g}$ to be recaptured) and $f_r$ (the probability for an atom left in $\ket{r}$ to be lost).

The ground state detection fidelity $f_g$ for a given sequence in which the tweezers are turned off for time $\tau$ is estimated by leaving the Rydberg lasers off and measuring the loss probability due to just the trap-off period. For short trap-off times $\tau < 4~\mu$s, the loss probability is $<1 \%$ so that $f_g > 0.99$. Longer trap-off times reduce this fidelity.

The Rydberg detection fidelity $f_r$ is characterized by the assumption that our single atom Rabi oscillation contrast is limited only by detection fidelity and finite coherence. This assumption is reasonable due to independent measurements which confirm our state preparation fidelity to be $>99.5\%$. From our fitted Rabi oscillations, we extract the amplitude at time $t=0$ to estimate the maximum possible loss signal which is typically $96(1)\%$. We associate the remaining $4\%$ with detection error, such that the Rydberg detection fidelity is $f_r = 0.96(1)$. This is consistent with careful analysis of the loss mechanism described in \cite{AntoineCoherence2018SI}.

\subsection{Rydberg laser alignment onto atoms}
We ensure consistent, centered alignment of each Rydberg excitation laser by picking off a small portion of the beam as it is coming to a focus and redirecting it onto a reference CCD camera (see Figure 1b in the main text). We first identify the location on the CCD camera that corresponds to optimal alignment onto the atoms. This is done by systematically displacing the beam to several different positions, and at each position measuring the intensity on the atoms through a measurement of the light shift on the microwave transition from $\ket{5S_{1/2}, F=2, m_F=-2}$ to $\ket{5S_{1/2}, F=1, m_F=-1}$. We fit these measurements, along with their corresponding positions on the CCD, to extract the optimal position. This procedure takes 5 minutes. We then keep the beam aligned onto this position on the camera by feeding back to a computer controlled mirror located one focal length before the final focusing lens. This realignment takes $< 5$ seconds and is performed every 2 minutes. This alignment procedure is used on both beams.

\subsection{Numerical model for single atoms}
The numerical model is implemented using the Python package QuTiP \cite{QuTiPSI}. It includes the following three effects:
\begin{enumerate}
\item A static but random Doppler shift in each iteration of the experiment. At $10~\mu$K, in a counter-propagating beam configuration with wavelengths 420 nm and 1013 nm, the random Doppler shift follows a Gaussian distribution of width $2\pi \times 43.5$ kHz.
\item Off-resonant scattering from the intermediate state $\ket{e} = \ket{6P_{3/2}, F=3, m_F=-3}$, to which both $\ket{g}$ and $\ket{r}$ are off-resonantly coupled by the 420 nm and 1013 nm lasers, respectively. The single-atom Rabi frequencies are $(\Omega_B, \Omega_R) \simeq 2\pi \times (60, 40)$ MHz, and the intermediate detuning is $\Delta \simeq 2\pi \times 600$ MHz.
 	
  We experimentally probe the decay channels of $\ket{e}$ by preparing atoms in $\ket{g}$ and then applying only the 420 nm light for varying amounts of time. After the 420 nm light is applied at the usual power and detuning, we perform microwave spectroscopy between the $\ket{F=1}$ and $\ket{F=2}$ ground state manifolds to determine the final atomic populations. We estimate that population leaves the Zeeman sublevel $\ket{g} = \ket{5S_{1/2}, F=2, m_F=-2}$ with a $1/e$ timescale of $\sim 150~\mu$s. We further estimate that population enters the $\ket{F=1}$ manifold on a timescale of $\sim 600~\mu$s.

  The optical scattering rate induced by the blue laser at the known detuning and power, along with the known $6P_{3/2}$ lifetime of 115 ns, leads to an estimated scattering timescale of $45~\mu$s, which is significantly faster than the timescale at which population leaves $\ket{g}$. The dominant decay channel from $\ket{e}$, then, is back into $\ket{g}$.

We therefore make the simplification in the numerical model that scattering events from $\ket{e}$ return population to $\ket{g}$. This process is modeled by Lindblad operators $\sqrt{\gamma_B} \ket{g}\bra{g}$ and $\sqrt{\gamma_R} \ket{g}\bra{r}$, corresponding to scattering events from the ground state $\ket{g}$ or the Rydberg state $\ket{r}$. The simulated timescales are $\gamma_B = 1/(40~\mu$s$)$ and $\gamma_R = 1/(80~\mu$s$)$.

\item Finite lifetime of the Rydberg state $\ket{r}$. The total effective lifetime of $146~\mu$s is composed of two decay channels: (1) blackbody stimulated transitions at 300 K to neighboring Rydberg states at a timescale of $230~\mu$s and (2) radiative decay of the Rydberg state to low principal quantum number levels at a timescale of $410~\mu$s \cite{RydbergProperties2009SI}.

  The simplified numerical model neglects radiative decay and only includes blackbody stimulated transitions to a new Rydberg state, $\ket{r'}$, which is dark to the laser field.
  
\end{enumerate}

\subsection{Definition of $\ket{W}$ state}
For experiments with two interacting atoms (as described in Figures 3 and 4 in the main text), the two atoms are coupled homogeneously by the same laser field to the Rydberg state. Strictly speaking, the Hamiltonian is given by
$$
\mathcal{H} = \frac{\hbar}{2} \sum_{i = 1,2} \left[\Omega e^{ikx_i} \ket{r_i}\bra{g_i} + h.c \right] + U\ket{rr}\bra{rr}
$$
The parameter $\Omega$ (taken to be real) is fixed by the laser intensities (which we assume here to be stable), and the parameter $k$ is fixed by the combined wavevector of the two counter-propagating lasers. The parameters $x_i$ describe the position of the two atoms along the array axis. In the Rydberg blockade regime where $U/\hbar \gg \Omega$, we project out the doubly excited state $\ket{rr}$ such that the only allowed levels are $\ket{gg}, \ket{gr}$, and $\ket{rg}$. We therefore rewrite the Hamiltonian as 

\begin{align}
  \mathcal{H} &= \frac{\hbar \Omega}{2} \left( e^{ikx_1} \ket{rg}\bra{gg} + e^{ikx_2} \ket{gr}\bra{gg} + h.c \right) \\
                      &= \frac{\hbar \sqrt{2}\Omega}{2} \left( \ket{W}\bra{gg} + \ket{gg}\bra{W} \right)
\end{align}
where we have defined
\begin{align}
  \ket{W} = \frac{1}{\sqrt{2}}\left(e^{ikx_1} \ket{rg} + e^{ikx_2} \ket{gr}\right)
\end{align}

In the main text, we work in the rotating frame in which the excited state $\ket{r}$ incorporates these position-dependent phase factors. Strictly speaking, however, the definition of $\ket{W}$ depends on the position of the atoms.

At the finite atomic temperature of $10~\mu$K, the atomic position along the array has a random Gaussian distribution of width $\sim 200$ nm. The relative phase $e^{ik(x_2-x_1)}$ may therefore reach appreciable values. We choose to not include these factors in our calculation of entangled state fidelity since they are fixed relative to the excitation laser system and do not emerge in measurements. Moreover, additional pulses that map Rydberg excitations down to other ground states may effectively erase these phase factors \cite{AntoineParity2014SI}. 

\subsection{Extracting off-diagonal matrix elements of density operator}
We consider an initial two-atom state $\rho_0$ that has the same measured populations as $\ket{W} = \frac{1}{\sqrt{2}}(\ket{gr} + \ket{rg})$ but unknown off-diagonal elements. Then $\rho_0$ can in general be expressed as
\begin{equation}
  \label{rho0}
\rho_0 = \frac{1}{2}(\ket{gr}\bra{gr} + \ket{rg}\bra{rg}) + (\alpha\ket{gr}\bra{rg} + h.c.)
\end{equation}
with off-diagonal coherence $|\alpha| \le 1/2$. We aim to measure $\alpha$, and in doing so to measure the entanglement fidelity of the two-atom system.

We consider the following protocol to determine $\alpha$. First, we apply a local phase shift operation that acts only on the left atom. This is achieved by a tightly focused laser which introduces a light shift on the ground state of the left atom. In the presence of this laser, $\ket{gr}$ is shifted by $\delta$ whereas $\ket{rg}$ is unshifted. After time $t$, the two states have accumulated dynamical phases $\ket{gr} \to e^{i\delta t}\ket{gr}$ and $\ket{rg} \to \ket{rg}$.

This operation transforms the density matrix from $\rho_0$  to $\rho_\phi$ where
\begin{equation}
  \label{eq:phase_shifted}
  \rho_\phi = \frac{1}{2}\left(\ket{gr}\bra{gr} + \ket{rg}\bra{rg}\right) + (\alpha e^{i\delta t} \ket{gr}\bra{rg} + h.c.)
\end{equation}
Note that if the state $\rho_0$ is a statistical mixture of $\ket{gr}$ and $\ket{rg}$ (that is, $\alpha = 0$), this phase shift operation does not change the density matrix.

We can rewrite $\rho_\phi$ from eq. \eqref{eq:phase_shifted} in the basis of the symmetric state $\ket{W} = \frac{1}{\sqrt{2}}(\ket{gr} + \ket{rg})$ and the orthogonal state $\ket{D} = \frac{1}{\sqrt{2}}(\ket{gr} - \ket{rg})$:
\begin{equation}
  \rho_\phi = \left(\frac{1}{2} + \alpha \cos (\delta t)\right) \ket{W}\bra{W} + \left(\frac{1}{2} - \alpha \cos(\delta t) \right) \ket{D}\bra{D} + \left[\left(-i \alpha \sin (\delta t)\right) \ket{W}\bra{D} + h.c. \right]
\end{equation}
Finally, a global resonant $\pi$-pulse at the enhanced Rabi frequency $\sqrt{2}\Omega$ maps $\ket{W} \to \ket{gg}$. The probability to end in $\ket{gg}$ is therefore the probability to be in $\ket{W}$ after the phase shift operation. Therefore
\begin{equation}
  P_{gg}(t) = \frac{1}{2} + \alpha \cos(\delta t) = 2\alpha \cos^2(\delta t/2) + \left(\frac{1}{2}-\alpha\right)
\end{equation}
The amplitude of the oscillation of $P_{gg}(t)$ as a function of $t$ therefore provides a direct measurement of $\alpha$.

While this derivation holds only for an initial density matrix of the form given in eq. \eqref{rho0}, a more general result can be found by considering the unitaries associated with the local phase shift, $Z_\phi^{(1)}$, and the $\pi$-pulse, $X_{\pi}^W$. In the basis $\ket{gg}, \ket{gr}, \ket{rg}, \ket{rr}$, the operators are given by:
\begin{align}
  Z_\phi^{(1)} =
  \begin{pmatrix}
	e^{i\delta t} & 0 & 0 & 0 \\
	0 & e^{i \delta t} & 0 & 0 \\
	0 & 0 & 1 & 0 \\
	0 & 0 & 0 & 1
  \end{pmatrix} \quad \quad
  X_\pi^W = 
  \begin{pmatrix}
	0 & \frac{i}{\sqrt{2}} & \frac{i}{\sqrt{2}} & 0 \\
	\frac{i}{\sqrt{2}} & \frac{1}{2} & -\frac{1}{2} & 0 \\
	\frac{i}{\sqrt{2}} & -\frac{1}{2} & \frac{1}{2} & 0 \\
	0 & 0 & 0 & 1
  \end{pmatrix}
\end{align}
For an arbitrary initial density matrix $\rho$, the final density matrix is $\rho' = U \rho U^\dagger$ where $U$ is the combined unitary $U = X_\pi^W Z_{\phi}^{(1)}$. The final probability to measure $\ket{gg}$ is then 
\begin{equation}
  P_{gg}(t) = \rho'_{gg, gg}= \frac{1}{2}\left(\rho_{gr,rg} e^{i\delta t} + \rho_{rg,gr} e^{-i\delta t}\right) + C
\end{equation}
where the constant $C$ is independent of the phase accumulation time $t$. Since $\rho_{gr,rg} = \rho_{rg,gr}^* = \alpha e^{i \theta}$ for some real amplitude $\alpha$ and angle $\theta$, we have
\begin{equation}
  P_{gg}(t) = \alpha \cos(\delta t + \theta) + C
\end{equation}
We therefore see that for any initial density matrix, the amplitude of the oscillation of $P_{gg}(t)$ as a function of $t$ gives a direct measurement of the off-diagonal coherence between $\ket{gr}$ and $\ket{rg}$.

\end{document}